\documentstyle[11pt,epsf]{article}

\def\fnote#1#2{\begingroup\def\thefootnote{#1}\footnote{#2}\addtocounter
{footnote}{-1}\endgroup}

\def\beq{\begin{eqnarray}}
\def\eeq{\end{eqnarray}}
\def\bea{\begin{eqnarray*}}
\def\eea{\end{eqnarray*}}

\def\NPB#1#2#3{Nucl. Phys. {\bf B#1}, #3 (#2)}
\def\PLB#1#2#3{Phys. Lett. {\bf B#1}, #3 (#2)}

\def\PRD#1#2#3{Phys. Rev. {\bf D#1}, #3 (#2)}

\def\PRL#1#2#3{Phys. Rev. Lett. {\bf #1}, #3 (#2)}

\def\centeron#1#2{{\setbox0=\hbox{#1}\setbox1=\hbox{#2}\ifdim
\wd1>\wd0\kern.5\wd1\kern-.5\wd0\fi
\copy0\kern-.5\wd0\kern-.5\wd1\copy1\ifdim\wd0>\wd1
\kern.5\wd0\kern-.5\wd1\fi}}
\def\ltap{\;\centeron{\raise.35ex\hbox{$<$}}{\lower.65ex\hbox{$\sim$}}\;}
\def\gtap{\;\centeron{\raise.35ex\hbox{$>$}}{\lower.65ex\hbox{$\sim$}}\;}

\def\lsim{\mathrel{\ltap}}

\def\singleandthirdspaced{\baselineskip=\normalbaselineskip\multiply
    \baselineskip by 130\divide\baselineskip by 100}
\def\singlespaced{\baselineskip=\normalbaselineskip}

\newcommand{\newc}{\newcommand}
\newc{\qbar}{{\overline q}}
\newc{\Kahler}{K\"ahler }
\newc{\deltaGS}{\delta_{\rm GS}}
\begin{document}
\begin{titlepage}
\begin{flushright}
{\large
hep-ph/9803432 \\
SLAC-PUB-7770 \\
SCIPP-98/09
}
\end{flushright}

\vskip 1.2cm

\begin{center}

{\LARGE\bf Dynamical supersymmetry breaking in models}

{\LARGE\bf with a Green-Schwarz mechanism}

\vskip 1.4cm

{\large
Nima Arkani-Hamed$^a$\fnote{\dagger}{Work supported by the Department of
Energy under contract DE-AC03-76SF00515. }, 
} 
Michael Dine$^{b,c}$, and  Stephen P.~Martin$^{b,c}$\\
\vskip 0.4cm
{\it $^a$Stanford Linear Accelerator Center,
     Stanford CA 94309} \\
{\it $^b$Santa Cruz Institute for Particle Physics,
     Santa Cruz CA 95064  } \\
{\it $^c$Physics Department,
     University of California,
     Santa Cruz CA 95064  } \\

\vskip 4pt

\vskip 1.5cm

\begin{abstract} 
We consider supersymmetry breaking in theories with gaugino condensation
in the presence of an anomalous $U(1)$ symmetry 
with anomaly cancellation by the
Green-Schwarz mechanism. In these models, a Fayet-Iliopoulos $D$-term
can give important contributions to the soft supersymmetry-breaking
scalar masses. Most discussions of this possibility have ignored the
dilaton field. We argue that this is not appropriate in general,
and show that the $F$-term contributions to the soft
breaking terms are comparable to or much larger than the
$D$-term contributions, depending on how the dilaton is stabilized.
We discuss phenomenological implications of these results.

\end{abstract}

\end{center}

\vskip 1.0 cm

\end{titlepage}
\setcounter{footnote}{0}
\setcounter{page}{2}
\setcounter{section}{0}
\setcounter{subsection}{0}
\setcounter{subsubsection}{0}

\singleandthirdspaced

When considering supersymmetry breaking, one of the most serious issues is
understanding the flavor structure of the soft supersymmetry breaking mass
terms.  There are several proposals to explain how a structure consistent
with known facts about flavor violation might arise:
\begin{itemize}
\item[1.] {\it Universal soft scalar masses at some high energy scale.} 
In the
context of, say, a supergravity theory, such a proposal is a convenient
starting point for phenomenology but is not, by itself, natural. It
corresponds to arbitrarily imposing a relation among a very large number
of parameters.  
\item[2.] {\it Dilaton domination.}  In string theory, if the
$F$ term of the dilaton is the principle source of supersymmetry breaking,
this leads to universal soft masses, provided one assumes that the \Kahler
potential for the dilaton is well-approximated by its weak coupling form. 
However, it is hard to understand how the dilaton potential can be
stabilized unless there are large corrections to the \Kahler potential. 
\item[3.] {\it Flavor symmetries.}  It is possible that approximate flavor
symmetries can give squark and slepton degeneracy or alignment, while
permitting the observed flavor violations among the fermions.  
\item[4.] {\it
Low energy, gauge mediated supersymmetry breaking.}  In such schemes,
gauge
interactions serve as the principle messengers of supersymmetry breaking. 
Soft breaking masses are functions of gauge quantum numbers, providing
adequate degeneracy to suppress flavor changing processes.  
\end{itemize}
The focus of this letter is a fifth suggestion: 
\begin{itemize} 
\item[5.] {\it Fayet-Iliopoulos $D$-term breaking as the source
of soft scalar masses.}  This possibility has been widely discussed in the
literature.  As proposed in \cite{BinetruyDudas} and \cite{DvaliPomarol},
this idea relies on the existence of a $U(1)_X$ gauge symmetry with
anomaly cancellation implemented by a non-trivial transformation of the
dilaton according to the Green-Schwarz mechanism \cite{GreenSchwarz}. 
After supersymmetry breaking and spontaneous breaking of $U(1)_X$, the
corresponding $D$-term obtains a vacuum expectation value (VEV). 
Light fields carry a $U(1)_X$
charge, so the $D$-term VEV contributes to
the soft squared masses of the squarks, sleptons and Higgs bosons of the
Minimal Supersymmetric Standard Model (MSSM).
If all of the squarks and sleptons carry the same $U(1)_X$ charge,
this can lead to flavor-independent contributions to soft
breakings. 
Alternatively, it might lead to interesting patterns of alignment, if the
Yukawa couplings are correlated with the $U(1)$ charges in just the
right way. These possibilities have been explored in recent
model-building, including \cite{IbanezRoss}-\cite{FaraggiPati}
\end{itemize}

This last proposal seems quite exciting.  It seems to
relate a very microscopic phenomenon (the generation of
a Fayet-Iliopoulos $D$-term through the Green-Schwarz mechanism) in a
quite well-defined and controllable
way to measurable properties of the low energy theory.  But
upon further consideration, the suggestion raises
several puzzles.  

First, one might wonder why $D$-terms should appear in the low energy
theory, given that the $U(1)_X$ gauge symmetry is broken at a very
high energy scale (one or two orders of magnitude below the Planck
scale), well above the scale of supersymmetry breaking.
Indeed, these terms can be understood from a low energy viewpoint
as arising from 
integrating out the corresponding massive vector supermultiplet.
This gives rise to corrections to the \Kahler potential for the
light fields, which in turn contribute to the low energy soft breakings.
To see this, consider first a general model with canonical \Kahler potential
terms for some chiral superfields $\phi_i$ with $U(1)_X$ charges $q_i$.
Then the scalar potential is
\beq
V = \sum_i \left | {\partial W \over \partial \phi_i}\right |^2 
-{1\over 2 g_X^2} D_X^2 - D_X (\sum_i q_i | \phi_i |^2+ \xi^2),
\eeq
where $W$ is the superpotential and $\xi^2$ is a constant Fayet-Iliopoulos
term in the lagrangian before symmetry breaking.
In the low-energy theory there are contributions
to the soft masses of the light fields
arising from the expectation value of the $D$-term:
\beq
m_{\phi_i}^2 = - q_i \langle D_X \rangle.
\label{mphisq}
\eeq
To relate this to properties of the light fields, note first
that
at a stationary point of the potential
the VEV of the $D$-term is related to
the $F$ term VEVs according to
\beq
\langle D_X  \rangle = -{g_X^2 \over M_X^2} \sum_i 
q_i |\langle F_i \rangle
|^2
\label{naiveDX}
\eeq
where $M_X^2 = g_X^2 \sum_i q_i^2 |\langle \phi_i
\rangle |^2$ is the (mass)$^2$
of the
$U(1)_X$ massive supermultiplet;
this is easily shown using the gauge invariance of $W$.
This corresponds to the fact that tree-level exchange of the heavy
gauge multiplet gives a contribution to the low-energy \Kahler
potential:
\beq
\Delta K = -
{g_X^2 \over M_X^2} q_i q_j \phi^{*i} \phi_i \phi^{*j} \phi_j.
\eeq
Now if we suppose that some of the $\phi_i$'s have non-zero $F$
components, eq.~(\ref{mphisq}) again follows.

Thinking about the problem in this way makes clear why one might
hope that the $D$ term provides the dominant contribution
to supersymmetry breaking. 
Since $U(1)_X$ is broken at scale one or two orders of 
magnitude lower than $M_P$,
the $U(1)_X$ gauge boson mass is lighter than the Planck scale
and so 
the (controllable) $D_X$ contributions to the soft masses 
$\sim 1/M_X^2$ can dominate over the (uncontrollable) $\sim 1/M_P^2$ 
supergravity
contributions.  

This way of thinking about the $D$ term suggests a strategy for
model building with $U(1)_X$ serving
as a ``messenger"
of supersymmetry  breaking.  One can consider a theory with
a sector which breaks supersymmetry, such as the $(3,2)$ model \cite{ADS},
and gauge a $U(1)$ symmetry.  The $F$ terms in the
symmetry breaking sector then give rise to a modification
of the \Kahler potential for any other fields charged under the
symmetry as in eq.~(4).
The resulting pattern of symmetry breaking then
depends on the charge assignments of the fields, and can produce
interesting patterns of degeneracy or alignment.   
Such models, however, 
suffer from some phenomenological difficulties. Scalars not charged 
under $U(1)_X$ (which typically include the top squark)
can only get soft masses from $1/M_P$ effects and are 
therefore much lighter than the charged scalars. 
Because
there are typically no low dimension, gauge invariant operators
in theories of dynamical supersymmetry breaking,
gaugino masses tend to be even further suppressed.\footnote{An exception
occurs in models with singlets.  However, in all such models, it
is necessary to prohibit some couplings.  This can only be done
naturally by imposing symmetries, which invariably suppress gaugino
masses.}
Apart from the usual
fine-tuning problem with heavy scalars \cite{DG}, this spectrum also
typically drives the top squark squared mass negative at the weak
scale
\cite{AM}.   

In the above scenario, the anomalous $U(1)$ is not in itself involved in the
dynamics of supersymmetry breaking. A more interesting possibility 
\cite{BinetruyDudas,DvaliPomarol} has $U(1)_X$ playing a crucial role in
the supersymmetry breaking dynamics. 
Schematically, these
models typically include a Standard Model singlet field 
$\varphi$, whose VEV and charge are appropriate to cancel the
Fayet-Iliopoulos term.  This field couples
to some fields charged under a non-abelian group. 
Upon integrating out these fields, gaugino 
condensation in the low energy theory generates a dynamical superpotential
for $\varphi$. The VEV of $\varphi$ needed to 
cancel the Fayet-Iliopoulos term is not at a stationary point 
of the superpotential, so supersymmetry is claimed to be broken. 

In these models, it is customary to ignore the dilaton (essential
for anomaly cancellation)
and assume that it does not play an important role
in supersymmetry breaking.  But analyzed in this way,
there is a puzzle:
a massless goldstino does
not appear in the spectrum.
The $U(1)$ gaugino and the fermionic component of $\varphi$ acquire a
Dirac mass from the Higgs mechanism, and there is no
light fermion arising from the non-abelian
dynamics.
The absence of a goldstino is clearly connected
with the anomaly.
To see this,
rather than considering the formal proof of the supersymmetric analog
of Goldstone's theorem, consider instead the explicit realization of the
theorem in weakly coupled theories with a canonical
\Kahler potential for all of the chiral fields.
Then the fermion mass matrix has
the form
\beq
M_{\rm fermion} = 
\pmatrix{
0   &  
\sqrt{2} g_X q_i  \phi^{*i}  \cr
\sqrt{2} g_X q_j \phi^{*j}  & 
{\partial^2 W \over \partial \phi_i \partial \phi_j}
}
\eeq
in the (gaugino, chiral fermion) basis with canonical kinetic terms.
Using the extremization condition for the scalar potential $\partial
V/\partial \phi_i = 0$, one finds that this matrix annihilates the 
eigenvector $\left (
\matrix{\langle D_X \rangle/\sqrt{2},\> g_X \langle F_i \rangle} \right )$
corresponding to the goldstino wavefunction, but {\it only} if the
superpotential is gauge invariant. 

This immediately resolves the puzzle of the missing goldstino in these
models.
In these theories, the superpotential $W$ is generally 
not gauge invariant unless
one maintains its explicit dependence on the dilaton  chiral
superfield $S$, which transforms under a $U(1)_X$ gauge transformation,
$A^X_\mu \rightarrow A^X_\mu + \partial_\mu \alpha$, according to
\beq
S \rightarrow S + i {\deltaGS \over 2} \alpha,
\eeq
with
\beq
\deltaGS = {1\over 192 \pi^2} \sum_i q_i.
\eeq
A typical non-perturbative superpotential has the form:
\beq
W_{\rm np} = \phi^A e^{-p S/\deltaGS}
\label{Wnp}
\eeq
where $p$ is a model-dependent positive number of order 1,  
and gauge invariance requires that $\phi^A$ carries $U(1)_X$ charge $p/2$.
Now it is apparent that in order to properly describe spontaneous
supersymmetry breaking with a massless goldstino, it is mandatory to
include $S$ as a dynamical degree of freedom along with the matter
fields. Indeed, the light degrees of freedom left after supersymmetry
breaking and $U(1)_X$ breaking must include $S$, and the goldstino
is predominantly the dilatino (the fermionic component of $S$).
This means that the dilaton $F$-term, ($F_S$), plays a crucial role
in supersymmetry breaking and its contributions to the soft breaking
terms in the low-energy theory cannot be neglected. 

To understand this, one can consider the origin of the $D$-term in the 
low-energy theory with the heavy fields (including fields which transform
under the strongly-coupled part of the gauge group) integrated out.
Consider a model which includes a chiral superfield $\varphi$ with 
$U(1)_X$ charge $-1$. In order to be gauge invariant, the \Kahler
potential
for the dilaton must be a function of $S+S^* - \deltaGS X$, where $X$ is 
the vector superfield for $U(1)_X$, so that
\beq
K_{\rm tot} = e^{-2X} \varphi^* \varphi + K(S+S^*-\deltaGS X).
\eeq
The $U(1)_X$ $D$-term can now be written as
\beq
D_X = - g_X^2 (\xi^2 -|\varphi|^2 )
\eeq
where $\xi^2 = -\deltaGS K'/2 > 0$ is the Fayet-Iliopoulos term
and $g_X^2 = 2/k_X(S+S^*)$ with $k_X$ the Kac-Moody level for $U(1)_X$.
Now, at the minimum of the potential $|\varphi|^2$ obtains a VEV which
nearly cancels $\xi^2$.
To relate this to the dilaton $F$-term VEV, one can make a 
field redefinition, shifting
$X \rightarrow X + {\rm ln} |\varphi|$ 
and $S \rightarrow S + {\deltaGS \over 2}~ {\rm ln} \varphi$ 
to obtain the ``unitary gauge" version of the \Kahler
potential:
\beq
K_{\rm tot} = e^{-2X} + K(S+S^* - \deltaGS X)
\eeq
in which the dependence on the absorbed field $\varphi$ has been eliminated.
Now one can integrate out the massive vector supermultiplet $X$ using its
equation of motion
\beq
X = -{1\over 2} {\rm ln}(-\deltaGS K') .
\eeq
(Here and in the following, a prime always means a derivative with respect
to $S$.) Taking the $D$-term component of both sides yields
\beq
D_X = -| F_S |^2 ({\rm ln} K')'' + {\deltaGS \over 2} D_X ({\rm ln} K')'
\eeq
so that at the minimum of the potential
\beq
\langle D_X \rangle  = |\langle F_S \rangle |^2 
\left ( - {K''' \over K'} + \left (
{K'' \over K'}\right )^2 \right ) \left (1- {\deltaGS K'' \over 2 K'}
\right )^{-1} .
\label{DXFS}
\eeq
Here $K$ is now taken to be a function of the scalar component of
$\langle S+S^* \rangle$. This general formula relates the $U(1)_X$
$D$-term to
the dilaton $F$-term VEV and the derivatives of the \Kahler potential.
The latter are constrained, but not determined, by the requirement
that the scalar potential is stable with respect to variations of the
dilaton.

If one were now to naively substitute the weak coupling form of
the \Kahler potential, $K = - {\rm ln} (S+S^*)$, into eq.~(\ref{DXFS}),
then one might conclude that $\langle D_X \rangle  = -
|\langle F_S \rangle |^2/\langle S+S^* \rangle^2$ up to small
corrections of order $\deltaGS/2\langle S+S^* \rangle$.
This already suggests that the $D$ term is not more important
than other contributions to the soft breakings.  However, this
\Kahler potential is only appropriate for large $S$, but the
true vacuum probably lies in a region where weak coupling
is not valid.  At the true minimum, this estimate may not 
be correct. For instance, if the dilaton is stabilized by 
\Kahler potential corrections, the derivatives
of $K$ cannot all be close to the weakly coupled prediction. To see this,
note that the dilaton-dependent part of the scalar potential includes
\beq
V = |W_{\rm np}' |^2/K'' + \ldots
\label{Vnp}
\eeq 
where $W_{\rm np}$ is of the form given in
eq.~(\ref{Wnp}). If this term dominates the contributions to the
minimization condition $V'=0$, then it follows that
\beq
{p\over \deltaGS} K'' = - K''' + \ldots
\label{Kppsuppression}
\eeq
so that $K''$ must be parametrically suppressed by one power of
$\deltaGS$ compared to $K'''$ at the minimum of the potential.
Of course if the dilaton is stabilized by corrections to the 
superpotential, the weak coupling estimate for $\langle D_X
\rangle $ given above
can be correct, so that $\langle D_X \rangle $ and $\langle F_S \rangle$
are comparable in size.

Despite recent progress in string theory, the mechanism
for stabilizing the dilaton -- if one exists -- is not known.
Various models have been proposed, including specially constructed
superpotentials \cite{racetrack}, and \Kahler potentials
motivated by non-perturbative string theory
considerations \cite{Shenker}-\cite{BCC}. For illustrative purposes, we
also consider a \Kahler potential which has a different
structure but which for our present purposes contains the
essential features of the latter class of models.
Our model is quite simple,
with a \Kahler potential chosen to have the correct behavior at weak
coupling,
a small number of parameters, and a minimum
of the desired sort. 
We take
\beq
K = -{\rm ln}(S+S^*) - {2 s_0 \over  S+S^*} + {b + 4 s_0^2 \over 6
(S+S^*)^2}
\label{toyKah}
\eeq
where $s_0$ and $b$ are non-negative constants. For large $S \gg s_0$,
this agrees with the weak coupling result. Now
\beq
K'' = {(S+S^* - 2 s_0)^2 + b \over (S+S^*)^4} .
\eeq
For $b>0$, this is positive-definite, as required for sensible kinetic
terms. In the limit $b\rightarrow 0$, it has a zero at $S=s_0$. Therefore,
the scalar potential eq.~(\ref{Vnp}) diverges at $S=s_0$ for $b\rightarrow
0$. In that case it is clear that the exponentially-falling superpotential
pushes $S$ out to a local minimum just less than $s_0$. As long as
we suppose\footnote{This evidently entails a fine tuning.}
that
$b \lsim \delta_{\rm GS}^2/p^2$, there will be a stable local minimum
near $S = s_0 - \deltaGS/p$. At that minimum, the derivatives of the
\Kahler
potential are given by (to leading order in $\deltaGS$):
\beq
K' = -{1\over 6 s_0};\qquad
K'' = {\delta_{\rm GS}^2 \over 4 p^2 s_0^4};\qquad
K''' = -{\delta_{\rm GS} \over 4 p s_0^4}.
\eeq
So we see that $K'''$ and $K''$ are {\it both} parametrically
suppressed, by $\deltaGS$ and $\delta_{\rm GS}^2$ respectively. To leading
order in $\deltaGS$, we therefore find that
\beq
\langle D_X \rangle = - |\langle F_S \rangle |^2 {K''' \over K'}
= -{3 \deltaGS \over 2p s_0^3} |\langle F_S \rangle |^2  .
\eeq
In the general class of models where the dilaton is stabilized by
a near vanishing of $K''$, we conclude that 
$\langle D_X \rangle $ is parametrically suppressed by $\deltaGS$ 
relative to $\langle F_S \rangle$.

The \Kahler potential of eqn.~(\ref{toyKah}) should provide a useful
toy model for dilaton stabilization in contexts other than
that considered here.  If one supposes that supersymmetry
is hierarchically broken, there is an approximate moduli space,
and it is presumably appropriate, even at strong coupling,
to write an effective action for the light fields such as
the dilaton.  In the present context, however, some of the
fields we are including in the effective lagrangian 
beneath the Planck scale (namely the $U(1)_X$ gauge field) 
approach Planck scale masses at strong coupling, 
and it is not clear that including
them incorporates the correct dynamics. However, 
we believe the model above gives some qualitative 
indication of the
correct physics, and at any rate, since the dilaton
should ultimately be stabilized at moderate 
coupling, the 
gauge multiplet may be light enough to justifiably be included in 
the low energy theory.

If the dilaton is stabilized by some other means, it is possible
to imagine that the suppression of $\langle D_X \rangle$ compared to
$|\langle F_S \rangle |^2$ that we have just found does not hold, even
though
eq.~(\ref{Kppsuppression}) is satisfied. This could be the case for
example if $K'''$ is extremely large at some value of $S$, corresponding
to a sudden change in $K''$. However, the nicest thing one can say about
such a possibility is that it is not particularly appealing.
As already mentioned, it is also possible to imagine 
that an unspecified superpotential effect
stabilizes $S$. Even in that case, however, eq.~(\ref{DXFS}) implies
that $\langle F_S \rangle$ is at least comparable to $\langle D_X
\rangle $.

It is instructive to consider how the preceding discussion is realized in
a concrete example, treating the gaugino condensation in the microscopic
theory explicitly. We will consider the model proposed by Bin\'etruy and
Dudas in \cite{BinetruyDudas}, with gaugino condensation from a
gauged $SU(N_c)$ symmetry, and taking $N_f=1$ for
simplicity. In addition to
the $SU(N_c)$-singlet field $\varphi$
with $U(1)_X$ charge $-1$, there are
chiral superfields $Q$ and $\overline Q$ transforming under 
$SU(N_c) \times U(1)_X$ as $({\bf N_c},q)$ and $(\overline{{\bf N_c}},
\overline q)$ respectively. It is convenient to minimize the potential
along the $SU(N_c)$-flat direction using the canonically normalized
meson superfield $t = (2Q\overline Q)^{1/2}$, so 
that the \Kahler potential is 
$t^* t(e^{2 q X} + e^{2 \overline q X})/2 + \varphi^* \varphi e^{-2 X} +
K(S+S^* - \delta_{\rm GS} X)$. 
Then the scalar potential is given by 
\beq
V = {K''} |F_S|^2 + \left |{\partial W \over \partial\varphi}
\right |^2 + 
\left |{\partial W \over \partial t} \right |^2 + {1\over 2 g_X^2} D_X^2,
\label{BDV}
\eeq
where $F_S = - W^{\prime *}/K''$ and
\beq
D_X = - g_X^2 ({q+\overline q \over 2} |t|^2 - |\varphi |^2 + \xi^2 ).
\eeq
The gauge-invariant superpotential is given by
\beq
W = m {t^2 \over 2} \left ( {\varphi \over M_P} \right )^{q+ \overline q}
+ (N_c-1) \left ( {2\Lambda^{3 N_c-1} \over t^2} \right )^{1\over N_c-1},
\eeq
where the last term is the ADS superpotential \cite{ADS} and corresponds
to $W_{\rm np}$ in eq.~(\ref{Wnp}). The dynamical scale $\Lambda$ depends
on the dilaton field according to
\beq
\left ( {\Lambda \over M_P }\right )^{3N_c-1} = 
e^{-8 \pi^2 k_N S} = e^{-2 (q+\overline q) S/\deltaGS}
\eeq
where $k_N$ is the Kac-Moody level of the $SU(N_c)$ gauge group.
The difference between the present treatment and \cite{BinetruyDudas}
is that we will include the effects of the first term $K'' |F_S|^2$
in the scalar potential eq.~(\ref{BDV}).

Now we can search for a local minimum of the scalar potential with
respect to
variations of $t$ and $\varphi$, in the neighborhood
of $\langle \varphi \rangle = \xi$. Following \cite{BinetruyDudas}, we can
define convenient parameters
\beq
\hat{m} \equiv m \left ( {\xi \over M_P} \right )^{q+\overline q};
\qquad\>\>\>\>\>
\epsilon = \left ( \Lambda \over \xi \right )^{3N_c-1 \over N_c}
\left (\xi\over \hat{m} \right )^{N_c-1 \over N_c}
\eeq
with $\epsilon \ll 1$.
Then the location of the minimum and the auxiliary field VEVs can be
determined as an expansion in $\epsilon$. One finds a local minimum at:
\beq
\langle \varphi^2 \rangle &=& \xi^2 \left [ 1 + \epsilon 
(q + \overline q) + \ldots \right ], 
\label{VEVphi}
\\
\langle t^2 \rangle &=& 2 \epsilon \xi^2 
\left [ 1 + \epsilon (q + \overline q)^2 {N_c-1 \over 2 N_c^2} 
\left (1 - 2 N_c - {2 K' \over \deltaGS K''} \right ) + \ldots
\right ]  .
\label{VEVt}
\eeq
At this minimum,
\beq
\langle D_X \rangle &=& {\epsilon^2 \hat{m}^2 (q+\overline q)^2 } 
\left [ 1 - {(q+ \overline q )\over N_c}
\left (1-{2K'\over \deltaGS K'' }
\right )
\right ] ,
\label{VEVDX}
\\
\langle F_S \rangle &=& \epsilon \hat{m} (q+\qbar ) {K' \over K''} ,
\label{VEVFS}
\eeq
to the lowest non-trivial order in $\epsilon$. (The $F$-terms for
$t$ and $\varphi$ also obtain VEVs, but they have a much smaller
effect on the soft masses of the fields in the low-energy theory.)
The terms in eqs.~(\ref{VEVphi})-(\ref{VEVDX}) which do not
explicitly involve $K''$ are the ones computed in \cite{BinetruyDudas}.
However, they are actually suppressed compared to the terms which
arise from including the $K'' |F_S|^2$ term in the potential,
at least for models where the dilaton is stabilized as 
described above. 

Using eqs.~(\ref{VEVDX}) and (\ref{VEVFS}), it is now possible
to compare the dominant sources of supersymmetry breaking:
\beq
\langle D_X \rangle/|\langle F_S \rangle|^2 &=& 
 {2(q+\qbar ) \over N_c \delta_{\rm GS}} {K'' \over K' } 
\left (1 - {\deltaGS K'' \over 2 K'} \right )
+ 
\left ({K'' \over K'} \right )^2 .
\label{DXFSmicro}
\eeq
At first sight, this does not seem to agree with eq.~(\ref{DXFS}),
since it does not even involve $K'''$. However, this is merely
because we have not yet used the minimization condition for the
dilaton, which in this model can be written as
\beq
{K''' \over K'} &=& 
- {2 (q+ \overline q) \over  N_c \deltaGS} {K''\over K'}
\left (1 - {\deltaGS K'' \over 2 K'} \right )^2 + 
{\deltaGS \over 2} \left ({K''\over K'}\right )^3
\label{Kpppmin}
\eeq
to leading order in $\epsilon$.
Using this one can show that eq.~(\ref{DXFSmicro})
is precisely equivalent to eq.~(\ref{DXFS}).
In particular, the $\deltaGS$ in the denominator of the RHS of
eq.~(\ref{DXFSmicro}) does {\it not} necessarily imply any enhancement of
$\langle D_X \rangle $; for example, in the model of dilaton stabilization
we discussed above, $K''$ is expected to be parametrically suppressed
by $\delta_{\rm GS}^2$.

As we discussed earlier, it is also possible to understand the presence
of the $D$-term in the low energy theory as arising from integrating
out the massive $U(1)_X$ vector supermultiplet. In a general model
with canonically normalized matter fields and a dilaton, a similar
argument to the one described earlier reveals that
\beq
\langle D_X \rangle = {g_X^2 \over M_X^2} 
\left ( {\deltaGS \over 2} K''' |\langle F_S \rangle |^2 - 
\sum_i q_i |\langle F_i \rangle |^2 - 
{\deltaGS\over 8} k_X \langle D_X \rangle^2 \right ).
\eeq
where now $M_X^2 = g_X^2 (\sum_i q_i^2 |\langle \phi_i \rangle |^2 +
\delta_{\rm GS}^2 K''/4)$.
In the model at hand with $\phi_i = \varphi,t$, this can be
checked to be in precise
agreement with eqs.~(\ref{DXFS}) and (\ref{DXFSmicro}), to
the lowest non-trivial order in $\epsilon$, by plugging in
the VEVs and
using
eq.~(\ref{Kpppmin}).

It is also easy to understand the emergence of the dilatino as the
goldstino in the microscopic picture. For a general theory with
a Green-Schwarz $U(1)_X$ symmetry and
chiral superfields $\phi_i$ with canonical \Kahler potential terms,
the fermion mass matrix is given by
\beq
M_{\rm fermion} = 
\pmatrix{ 
{k_X g_X^2 W^{\prime *}/2 K^{\prime\prime}} &
\sqrt{2} g_X q_i \phi_i^* & 
{g_X \over \sqrt{2}} (
{k_X D_X \over 2 \sqrt{K^{\prime\prime}} }
- 
\delta_{\rm GS} \sqrt{K^{\prime\prime}} 
)
\cr
\sqrt{2} g_X q_j  \phi_j^* & 
{\partial^2 W \over \partial \phi_i \partial \phi_j} & 
{1\over \sqrt{K^{\prime\prime}}} {\partial W' \over \partial \phi_j}
\cr
{g_X \over \sqrt{2}} (
{k_X D_X \over 2 \sqrt{K^{\prime\prime}} }
- 
\delta_{\rm GS} \sqrt{K^{\prime\prime}} 
)
&
{1\over \sqrt{K^{\prime\prime}}} {\partial W' \over \partial \phi_i}
&
{W^{\prime\prime} \over K^{\prime \prime}} -
{K^{\prime \prime \prime} W^{\prime } \over  K^{\prime \prime 2}}
\cr
}
\eeq
in the canonically-normalized (gaugino, chiral fermion, dilatino)
basis.
The goldstino wavefunction in this basis is proportional to
\beq
\widetilde G = \left (
{\langle D_X \rangle \over \sqrt{2} g_X}, \langle F_i
\rangle , \sqrt{K''} \langle F_S \rangle
\right ).
\eeq
The first row of $M_{\rm fermion}$ annihilates
$\widetilde G$ by virtue of 
the gauge invariance
condition
\beq
\sum_i q_i \phi_i {\partial W\over \partial \phi_i} 
- {\delta_{\rm GS} \over 2} W' = 0,
\eeq
while the second and third rows annihilate $\tilde G$
by the minimization conditions $\partial V/\partial \phi_i = 0$ and
$V' = 0$ respectively. 
Now, we can specialize to the model studied above. Looking only at the
lowest order contributions in $\epsilon$, one finds that in the basis
$(\lambda_X/g_X,\psi_t, \psi_\varphi, \psi_S/\sqrt{K^{\prime\prime}})$,
the goldstino wavefunction is proportional to
$(0,0,\sqrt{K''/K'},1)$.
(The zeros actually correspond to terms suppressed by $\epsilon$
and $\sqrt{\epsilon}$.) So in the scenario for dilaton stabilization
discussed
above, the goldstino is mainly dilatino with a small admixture of
the fermionic component of $\varphi$.

In the preceding discussion we have been using a global supersymmetry
picture. Including supergravity effects causes the gravitino to
obtain a mass by absorbing the goldstino, but does not alter the
essential features of the supersymmetry breaking pattern. In particular,
including the minimal supergravity terms in the scalar potential
does not affect the ratio $\langle D_X \rangle/|\langle F_S \rangle |^2$
to leading
order in $\deltaGS$.

Let us conclude by noting some phenomenological implications of
this analysis. Previously, it was thought that models 
of dynamical supersymmetry breaking 
in the presence of an anomalous $U(1)$ featured very small
gaugino masses and scalar squared masses dominated by the $D$-term VEV.
However, the picture that now emerges is similar to that of a
moduli-dominated
scenario \cite{dilatondominated}, but with small $D$-term corrections.
In the theory below $M_X$, assuming a canonical gauge
kinetic function (for a possible rationale for this,
see \cite{BanksDine}) the MSSM gauginos will
obtain masses
\beq
m_{\lambda} = {\langle F_S \rangle \over \langle S+S^* \rangle} .
\eeq
Each of the MSSM scalars with
$U(1)_X$ charge $q_i$ receives 
\beq
m^2_{\phi_i} =
{1\over 3} K'' |\langle F_S \rangle|^2
- q_i \langle D_X \rangle + \cdots
\label{scalarmass2}
\eeq
where the first term
represents the usual minimal contribution of the $F$-term 
of the dilaton, the second is the anomalous
$U(1)_X$ contribution, and the ellipses refer to other contributions to
the soft masses coming from higher order \Kahler potential couplings
between $\phi_i$ and $S$, and from contributions due to $F$-terms
of other moduli. It is important to remember that such contributions
can be comparable to the terms shown explicitly, and need not have any
special flavor structure.
It follows from eqs.~(\ref{VEVDX}) and (\ref{VEVFS}) that the
contributions to the soft
masses from $\langle F_S \rangle$ and $\langle D_X \rangle$ are both 
proportional to $\epsilon \hat{m}$, and so can be made
comparable to the electroweak scale by a natural choice of the dynamical
scale $\Lambda$.
Note that $\langle D_X \rangle$ turns out to be negative in our
conventions,
so that the $D$-term contributions to MSSM scalar squared masses are
positive for $q_i > 0$. 
We have found that the $D$-term contributions to scalar
masses are likely to be parametrically suppressed by $\deltaGS$ compared to
the $F$-term contributions to the gaugino masses (note that $\langle S
\rangle$ is typically of order 2
or so) in models where the dilaton is stabilized by large corrections
to $K$. If the $F-$term contribution to the scalars are suppressed, 
it is possible
that the $D$-term contributions dominate the
tree-level scalar masses. 
However, renormalization group running yields large
flavor-independent positive contributions to the scalar (mass)$^2$
proportional to $m_{\lambda}^2$, so that the physical masses of squarks
and sleptons are again not dominated by the $D$ term.

Since $D_X$ is not the dominant source of 
supersymmetry breaking, we cannot use the anomalous $U(1)$ as a
controllable
handle on the soft masses. 
If these theories are to have any hope of
being realistic, we must assume that
(a) none of the other moduli
acquire large $F$-component VEVs and (b) the 
contributions to scalar masses due to 
$\langle F_S \rangle \neq 0$
are flavor-blind. 
Even with these assumptions,
it is a quite model-dependent question whether these theories 
can be consistent with present phenomenological constraints on flavor
violation. If the $U(1)_X$ charges are family-independent, as in the
models in \cite{Ramond,FaraggiPati}, then we expect that the $\langle D_X
\rangle
$ contributions
are harmless for flavor-violation even though they are not universal;
this makes them particularly interesting for future sparticle
spectroscopy.
(Of course, other family-dependent $D$-terms present in
such models might very well still be
dangerous.) On the other hand, in models where the $U(1)_X$ symmetry
is family-dependent, there is a quite serious flavor-violation
problem unless the $D$-term contributions to the down squark and
slepton squared masses happen to be aligned with the 
corresponding fermion Yukawa
couplings. The presence of larger universal $F_S$-term
contributions
may well ameliorate this problem, and a rough estimate
shows that the relative suppression of the $D$-term contributions
$\sim \delta_{GS} \sim 10^{-2}-10^{-3}$ may just be sufficient
to explain the absence of flavor-changing neutral currents
(for 1 TeV squarks). 


If low energy supersymmetry does
have something to do with nature, the flavor problem is surely an
important clue as to how supersymmetry is broken.  If the breaking
is at a high scale, one might have hoped that $D$-term breaking
with an anomalous $U(1)$ could help resolve this problem. 
In theories where the $U(1)$ merely serves as a ``messenger" of 
supersymmetry breaking, this could indeed happen, although the gaugino
masses
tend to be very light and the sfermion spectrum has fine-tuning problems.
In theories where the anomalous $U(1)$ dynamics is involved in
supersymmetry breaking,
we have learned that contrary to the naive expectation, 
the $D$-term contributions to
soft terms in the low energy theory do not, in fact, dominate over Planck 
scale contributions.   
Therefore theories of this sort are still subject to potentially dangerous
flavor-violating effects from non-minimal contributions to the \Kahler
potential which involve both $S$ (and the moduli) and the light fields.
If we assume that such large flavor-violations are absent, however, 
anomalous $U(1)$ theories can still be useful for generating the fermion 
mass hierarchy while evading flavor-changing constraints.
There is another positive aspect of our observations. It is usually asserted
that in these models the gauginos tend to be very light; this is now
seen to be not the case.

In the introduction, we listed five mechanisms for resolving
the flavor problems of supersymmetric theories.   In this
paper we have asked in what sense the fifth, supersymmetry
breaking through $D$-terms, is special.  We have argued that
one should think about this mechanism by integrating out
the massive vector field(s). 
If the mechanism is to be effective, it is crucial that the
resulting terms dominate, i.e. that the vector masses be small
compared to, e.g. the Planck or string scale.  In such
a case, soft breaking masses will be controlled
by the $U(1)$ charges of the fields -- this is the
real significance of $D$ term breaking.  But we have
seen that in theories where the dilaton plays the dominant
role in supersymmetry breaking, the couplings of the dilaton
to the vector fields are suppressed.  	In theories such as the
$(3,2)$ model, when coupled to the $U(1)$, the $D$ term
can dominate the scalar soft breakings, but gaugino masses
and the $\mu$ term will be difficult to explain.

NAH would like to thank Lance Dixon for useful discussions,
and MD thanks Yossi Nir for several comments.
This work was supported in part by the US Department of Energy.


\end{document}